Clinical dichotomania: A major cause of over-diagnosis and over-treatment?

Huw Llewelyn MD FRCP
Department of Mathematics
Aberystwyth University
Penglais
Aberystwyth
SY23 3BZ
Tel 0044 7968528154

hul2@aber.ac.uk

Abstract

Introduction: There have been many warnings that inappropriate dichotomisation of results into positive or negative, high, or normal etc., during medical research could be very damaging. The aim of this paper is to argue that this is a major cause of over-diagnosis and over-treatment.

Methods: Illustrative data were taken from a randomised control trial (RCT) that compared the frequency of nephropathy within 2 years in those on treatment with an angiotensin receptor blocker and a control on patients in whom the numerical value of the albumin excretion rate (AER) was available on all patients before they are randomised.

Results: When the results of a RCT were divided into AER ranges, a negligible proportion of patients developed nephropathy within 2 years in the range 20 to 40mcg/min and therefore would be unlikely to accept treatment during shared decision making; 36% of currently treated patients fell into this range. Above an AER of 40mcg/min, there was a gradual increase in proportions with nephropathy in each range, with fewer developing nephropathy in each range on irbesartan 150mg daily than on control and fewer still developing nephropathy on 300mg daily.

Interpretation: When logistic regression functions were fitted to the data and calibrated, curves were created that allowed outcome probabilities and absolute risk reductions to be estimated for use in diagnosis, the offering of treatment and subsequent decision making. This could avoid much overdiagnosis and overtreatment caused by an overestimation of outcome probabilities.

Discussion: If this approach is applied to patients with other significant risk factors (e.g., HbA1c for nephropathy) then careful attention should be given to the principles of diagnostic reasoning and causal inference regarding additive and multiplicative scales to avoid over-estimating risks.

Conclusion: Careful attention to diagnostic severity and its effect on outcome probabilities by interpreting each numerical diagnostic result provides better application of the principles of diagnosis and treatment decisions that can reduce over-diagnosis and over-treatment.

Introduction

The late Doug Altman warned that inappropriate dichotomisation of results into positive or negative, high, or normal etc. during medical research could be very damaging [1]. This warning has also been given by many others and has been described as 'dichotomania' [2, 3]. It often happens when numerical test results that represent the severity of a condition are dichotomised into 'normal' when they are within two standard deviations of the mean of a population. or 'abnormal' if they are above or below two standard deviations of the mean. Such thresholds can be based on a variety of different populations [4]. The importance of applying the results of RCTs to the baseline probabilities of individual patients has been emphasised already (e.g., by assuming that the risk ratio (or risk reduction) is constant [5]). Unfortunately, it is the average risk difference (AKA the average absolute risk reduction) in a RCT based on a single threshold that is often applied, which overestimates many of the risk ratios or risk differences and under-estimates the others.

The aim of this paper is to argue that application of inappropriate thresholds for diagnostic and treatment indication criteria is a major cause of over-diagnosis and over-treatment by causing over-detection and over-definition [6]. It will be argued that this can be minimised by estimating the probability of an outcome at every degree of diagnostic severity on treatment and control in a RCT. Thresholds for treatment indication criteria and diagnostic criteria are then established based on the way these probabilities affect treatment decisions with or without shared decision making by well-informed people [7] with or without decision analysis [8, 9]. This is a different concept to that of over-diagnosis based on the outcome of not treating people with a diagnosis and waiting to see what proportion of patients come to no harm [6].

Methods

The example used to support the argument is the diagnosis and treatment of albuminuria in patients with diabetes mellitus. The illustrative data were taken from the IRMA2 randomised control trial (RCT) that compared the outcome of treatment with an angiotensin receptor blocker (ARB) and a control on patients with 'Diabetic Albuminuria' (DA) in those whose hypertension was controlled [10, 11]. In this analysis, 'albuminuria' was defined as the average albumin excretion rate (AER) of at least 20mcg/min when measured on 3 separate occasions. The samples were collected on rising (thus assuming no orthostatic nor exercise induced albuminuria), the MSU showed no excess white cells or bacteria or growth (thus assuming no infection) and with no red cells (thus assuming no nephritis). If all these conditions were met, a patient was assumed to have DA. The average of the three pre-randomisation AER values was used as a measure of assumed severity of the DA. The patients with this diagnosis of DA had been randomised to treatment with either placebo or irbesartan 150mg or 300mg daily, the blood pressure being then controlled within 3 months based on a target below 135mmHg systolic in all trial limbs by addition of agents that did not act via the renin angiotensin system. The outcome of 'nephropathy' was defined as all those and only those who exceeded an AER of 200mgc/min within 2 years and a 30% rise from the pre-randomisation AER.

Assumptions

It was assumed that a diagnosis is not the same as a disease. A diagnosis is an assumption that a disease is present and is therefore a theory. Diagnostic criteria are therefore rules for justifying making such an assumption for practical purposes. The practical purpose of a diagnosis (e.g., DA) is to identify the possible treatments, follow up strategies, causes and complications etc. that might apply to a patient. A treatment indication identifies the sub-group of those with a diagnosis to whom a treatment should be offered (which can therefore be regarded as more detailed sub-diagnosis). In some cases, the criterion for a diagnosis is the same as for the treatment indication, in which case all patients with the diagnosis are offered the treatment. The offered treatment is given only after the patient accepts it during decision making. The aim of the above methods is to explore how diagnostic and treatment indication criteria are established based on a decision. It is done by examining the probabilities of an outcome (i.e., nephropathy) at various degrees of diagnostic severity (i.e., at various values of AER) with and without treatment. The range of probabilities at which well-informed people begin to choose treatments is used to identify the threshold for a treatment indication criterion when treatment should be offered in future. If there are several interventions suggested by a diagnosis, then the lowest interventional or treatment threshold is adopted as the threshold for the diagnostic criterion.

Results

Table 1 shows the overall result of the IRMA 2 study. Based on grouping them into a single AER range of 20 to 200mcg/min, this would suggest that there is a 15.3% of developing nephropathy within 2 years on control, 10.2% on Irbesartan 150mg daily and 5.3% on Irbesartan 300mg daily. If the groups treated with irbesartan 150 and 300mg daily were combined, then 7.73% developed nephropathy.

Table 1: The proportions developing nephropathy within 2 years on of control versus Irbesartan 150mg daily and 300mg daily in a RCT and the average of Irbesartan ({150+300)/2} mg daily.

| AER range | Control limb | Irbesartan 150mg | Irbesartan 300mg | Irbesartan {(150+300)/2}mg |
|---|---|---|---|---|
| 20 to 200µg/min | 30/196 = 15.3% | 19/186 = 10.22% | 10/189 = 5.29% | 29/375 = 7.73% |

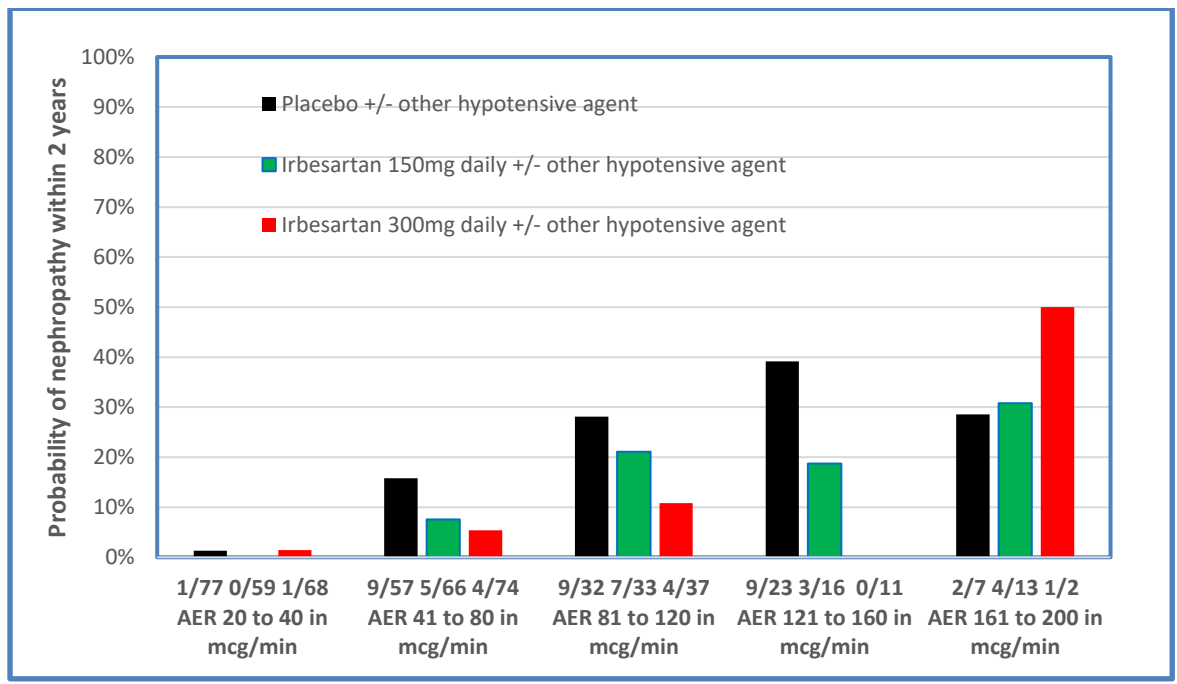

Figure 1: 'Stratifying' RCT results based on ranges of diagnostic severity of Diabetic Albuminuria (i.e., the albumin excretion rate in mcg/min)

Figure 1 displays the proportions developing nephropathy within 2 years at various ranges of AER on control, Irbesartan 150mg daily and Irbesartan 300mg daily. Table 2 shows the proportions of patients with nephropathy who had an AER up to 80mcg/min and over 80mcg/min in those on Control and those on Irbesartan 150 or 300 mg daily (i.e., an average dose of (150+300)/2 mg daily) obtained from the proportions in Figure 1.

Table 2: The proportions of patients with nephropathy who had an AER up to 80mcg/min and over 80mcg/min in those on Control and in the combined group of Irbesartan 150 or 300 mg daily.

| Intervention | AER ≤ 80mcg/min | AER > 80mcg/min |
|---|---|---|
| Control | 10/134 | 20/62 |
| Irbesartan 150 or 300mg daily | 10/264 | 19/111 |

Data analysis and interpretation

The methods here are examples of the ways in which the above data might be interpreted. Thus, to estimate what the results might be after an infinite number of observations, the results from irbesartan 150mg daily and 300mg daily were combined initially because of paucity of data on those with nephropathy on treatment. The proportions developing nephropathy at AER ranges 20 to 40mcg/min, 41 to 80, 81 to 120, 121 to 160 ad161 to 200mcg/min as shown in Figure 1 were used to calculate natural logs of the corresponding odds to construct logistic regression function for an average dose of Irbesartan ((150 +300)/2} mg daily. A similar curve was constructed for control.

The curves were calibrated by adjusting the coefficients of an additional linear function until the average probabilities read from both curves above and below 80mcg/min (using the data used to construct the curves) corresponded to the observed proportions above and below this threshold. The AER threshold of 80mcg/min was chosen because at this point the sensitivity roughly equalled the specificity. This 'calibration' was done with the 'training data' to ensure that the average of all the probabilities matched the overall frequencies in Tables 1 and 2. It is assumed that the logistic function avoids over-fitting and that calibrating with a linear function does not change this. The curves could be calibrated again with different test data, but the latter were not available.

The curves for Irbesartan 150mg daily and 300mgm daily were obtained by assuming that the distribution of the AERs and the likelihood distribution of AERs in those with nephropathy are the same in those on Irbesartan 150mg daily, {(150+300)/2}mg daily and 300mg daily, so that the ratio of the latter probability distributions at each AER value is the same on the three curves. This means that for any AER on the curve for 150mg daily the probability of nephropathy is equal to the probability of nephropathy on the (150+300)/2 curve divided by 29/375 and multiplied by 19/186 (see Table 2). Similarly, for any AER on the curve for 300mg daily, the probability of nephropathy is equal to the probability of nephropathy on the {(150+300)/2) mg curve divided by 29/375 and multiplied by 10/189 (see Table 2). The resulting curves for the control data and those for Irbesartan 150mg daily and 300 mg daily are shown in Figure 2.

The 95% confidence intervals were estimated based on the assumption that the variance of the distributions of AER results of those with and without nephropathy was minimal and so the variance of the values at each point of the curve was assumed to be approximately $P +/- 1.96*(P*(1-P)/N)^{1/2}$ when P was the probability of nephropathy read from the curve and N was the total number of observations used to construct the curve (N = 196 for the control curve, N = 186 for the Irbesartan

150mg daily and 189 for irbesartan 300mg daily). Figure 2 shows the result of fitting these curves for the 95% confidence limits. These curves that represent diagnostic severity are considered theoretical as they represent assumptions about the underlying disease process in keeping with the theoretical nature of a diagnosis in general (not hypothetical because they cannot be tested empirically by collecting an infinite amount of data). As theories, they are consistent with the observed frequencies of nephropathy in those on Irbesartan 150 mg daily and 300mg daily as shown in Tables 1 and 2.

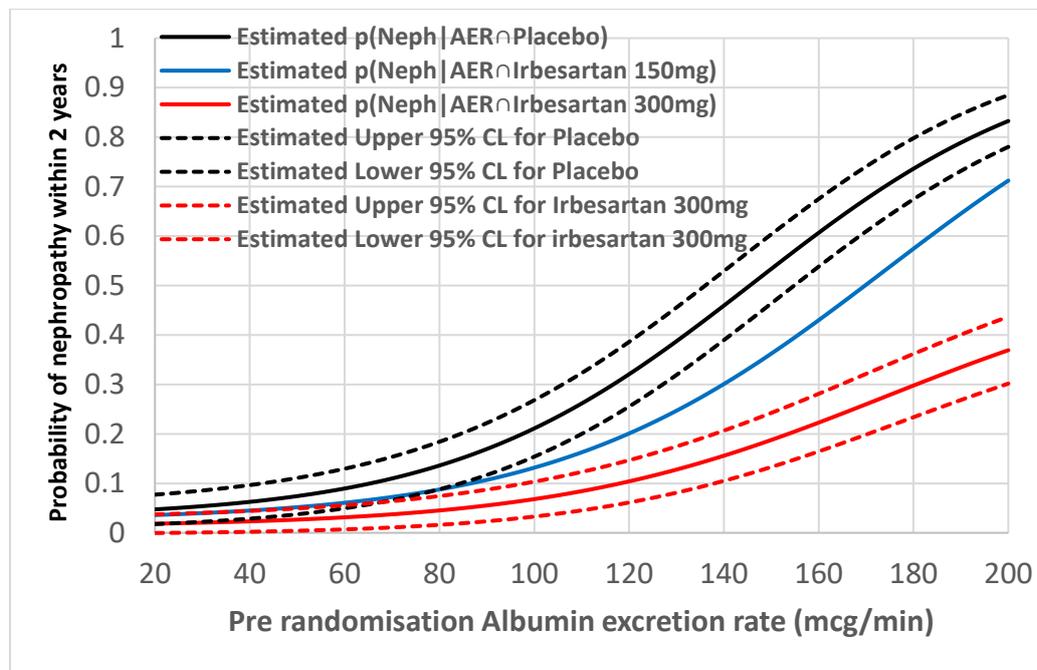

Figure 2: Probabilities of outcome due to severity of Diabetic Albuminuria represented by the albuminuria excretion rate in mcg/min.

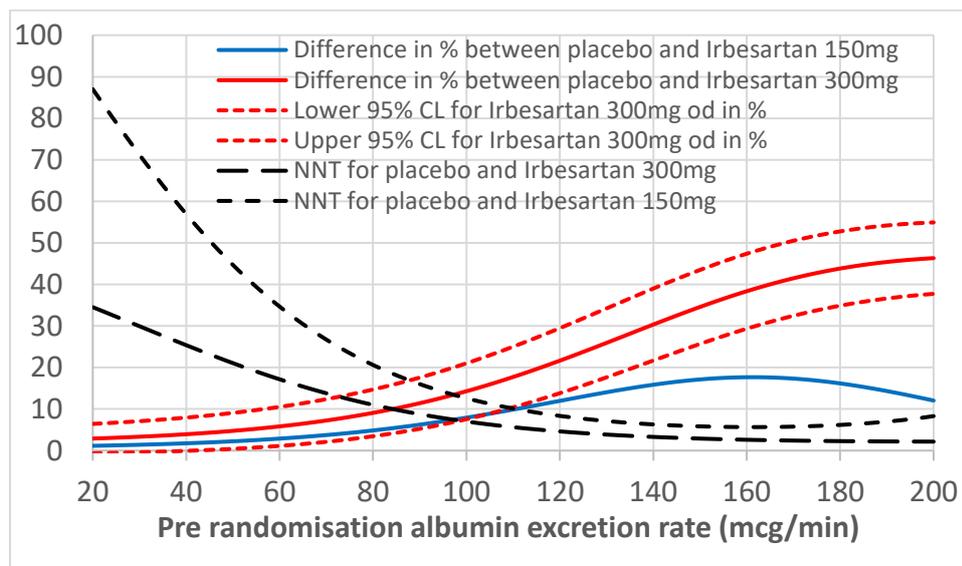

Figure 3: The estimated risk differences and the NNTs for Irbesartan 150 and 300mg daily at each albumin excretion rate with risk difference 95% confidence limits for Irbesartan 300mg daily Figure 3 near here

The risk differences a shown in Figure 3 were found by subtracting the probability of nephropathy conditional on irbesartan from its probability on control at each AER value. This was done for both doses of Irbesartan. They were expressed as percentages so that the NNT could be plotted on the same vertical scale. The number needed to treat was calculated for each AER as the reciprocal of the risk difference. The 95% confidence limits at each AER for Irbesartan 300mg daily were estimated as follows: $(P_c-P_t)+/-1.96((P_c(1-P_c)/196)+(P_t(1-P_t)/192))^{0.5}$ (when 'Pc' is the probability of nephropathy conditional on the relevant AER and control read from the curve, 'Pt' is the analogous probability of nephropathy conditional on treatment with irbesartan, 196 the number of observations used to construct the control curve, and 192 the number of observations used to construct the treatment curve. The estimated curves as shown in Figures 2 and 3 might represent the impression that an experienced physician would form mentally about changing outcome probabilities due to diagnostic severity from observing the raw results in Figure 1.

The 95% confidence intervals give an indication of the variation in probability of nephropathy in the curves if the RCT on which they were based was repeated many times. However, if the measurement of the AER in an individual patient to whom Figure 2 were applied were repeated many times the AER results would vary too. This variation could be estimated by calculating the standard deviation of the AER results of such repeat measurements if it could be assumed that an infinite number of such results were normally distributed. In that case, 95% of the results of such individual AER measurements would lie between +/- 1.96 standard deviations of the mean. This would be known as the 95% predictive interval for the AER. The probability of nephropathy based on the latter and the 95% confidence intervals in Figure 2 would therefore lie within a wider interval than the 95% confidence interval alone shown in Figure 2. However, the gradient of the curve in the lower AER regions is shallow, so the interval within which the probability of nephropathy read from the curve lies may not be very wide on the vertical scale displaying the probabilities of nephropathy.

Precision medicine that abolishes over-diagnosis and over-treatment

A perfect test (e.g., a new AER*) would provide a horizontal 'curve' at lower values indicating an outcome probability of zero (e.g., up to a position of an AER*of 40mcg/min). It would then rise vertically from that point to reach an outcome probability of one (certainty) at the position of an AER* of 41mcg/min. A perfect AER* would also have a profoundly powerful mechanistic link to the outcome, and have high measurement reproducibility, so that the relative effect of other randomly variable risk actors would be negligible by comparison. Therefore, the 95% confidence and predictive intervals would be extremely narrow. A perfect treatment would reduce the outcome probability to zero at an AER* of 41mcg/min and above and create no adverse effects. This would greatly simplify decisions – not to treat anyone with an AER* up to 40mcg/min and to treat everyone above this threshold. This would abolish over-diagnosis and over-treatment and is the dream of precision medicine. In the meantime, we can only try find tests using hypotheses based on knowledge and theories of disease to make the curves steeper with lower variation and to find treatments than lower the curve as much as possible without causing adverse effects.

Application during decision-making

If a patient with Type 2 Diabetes Mellitus and a systolic blood pressure of less than 135mmHg (with or without treatment with a non-ACE inhibitor or non-ARB) presented with an AER of 40mcg/min, she could be told that the probability of developing nephropathy within 2 years without treatment would be approximately 0.062. She could be given detailed pictures of what life would be like with this diagnosis. This picture might entail a referral to a renal clinic, more tests and a range of possible treatments and outcomes that included end-stage renal failure and dialysis. However, if she took Irbesartan 300mg daily then the probability developing nephropathy would be about 0 023 ('about' means bearing the 95% confidence and predictive intervals in mind). The risk difference would be about 0.062-0.023 = 0.039 so the number needed to treat (NNT) for one to avoid nephropathy would be about 1/0.023 = 25 (see Figure 3). (The 95% confidence intervals for the NNT would be found using the 95% confidence intervals for the risk difference.)

If the patient and or doctor considered the inconvenience, and possible adverse effects of taking Irbesartan then this might result in a decision not to treat. In practice, the decision would be made intuitively but many would advocate more formality by using decision analysis [8, 9]. However, the current approach would be to base the probability of nephropathy without treatment on Table 1 and suggest that it was 0.153, reducing to 0.053 on Irbesartan 300mg daily. This exaggeration might result in a decision to accept the treatment and result in 'over-treatment'. According to Figure 2, these probabilities of 0.153 and 0.053 would only be reached if the AER were 85mcg/min.

If a study was performed on volunteers who were asked to assume that there were no adverse effects from an ARB, then despite this, few people if any might accept treatment at an AER of 40mcg/min or below when the risk difference is no more than 0.039 and the NNT is at least 25. Thirty six percent of patients had an AER between 20 and 40mcg/min. Therefore, if the threshold for diagnosis was moved to 40mcg/min, it would mean that 36% of patients with an AER of at least 20mcg/min would previously have been over-diagnosed. Above an AER of 40mcg/min few people might still accept treatment if provided with the information in Figures 2 and 3. Perhaps the threshold for the diagnosis of 'DA' and thus considering its treatment might be moved even higher than 40mcg/min. At an AER of 60mcg/min, the risk difference is 0.090-0.031 = 0.059, the NNT is still 17 and 57% patients would have an AER between 20 and 60mcg/min. At an AER of 80mcg/min, the risk difference is 0.142-0.047=0.095, the NNT is 11 and 72% would have an AER between 20 and 80mcg/min suggesting that 72% of patients would have been over-diagnosed previously if the threshold was 20mcg/min. Note that these diagnostic thresholds would be based on the outcome of decisions. The implications of these thresholds are summarised in Table 3.

From Figure2, if treatment is rejected at an AER of 20mg/min or below, the probability of nephropathy would be no higher than 0.05 but at AER up to 80mg/min the probability of untreated nephropathy is no higher than 0.15. This represents the consequence of underdiagnosis and under-treatment. The latter probability might be acceptable during shared decision making if the outcome is not fatal and the test can be repeated, but not if the outcome was death due to cancer for example. Information of the kind in Table 3 thus allows a trade-off to be considered between possible over-treatment and under-treatment.

Table 3: Proportion of patients regarded as over-diagnosed in retrospect if a new threshold is adopted.

| New AER threshold adopted for the diagnosis and treatment of Diabetic Albuminuria | NNT at new threshold if it is adopted for treatment with Irbesartan 300mg daily | % previously over-treated in retrospect at a threshold of 20mcg/min if new threshold adopted | Maximum percentage with the outcome that is not treated (i.e., under-treated) if new threshold adopted |
|---|---|---|---|
| AER = 20mcg/min | 35 | 0% | 5% |
| AER = 40mcg/min | 25 | 36% | 6% |
| AER = 60mcg/min | 17 | 57% | 9% |
| AER = 80mcg/min | 11 | 72% | 15% |

Current practice

In practice, the above situation rarely arises. If people with diabetes mellitus need treatment for a high blood pressure, the treatment of choice is an ACE inhibitor or an ARB such as Irbesartan that would reduce any albuminuria before it could be detected. In other words, the above situation is pre-empted by prescribing an ACE inhibitor or ARB first to avoid the need to add it later. The main purpose of this example therefore is to illustrate the principles of avoiding arbitrary thresholds but basing them on choices during decision making. This is a general principle that should be applied to all numerical results and that numerical results for use as evidence to predict outcomes. The results should be interpreted on their own merits first and not dichotomised immediately. However, dichotomisation is necessary when an outcome is a diagnosis such as nephropathy that is linked to a provisional decision to consider or propose treatment before an actual decision is made. This dichotomisation for the purpose of considering a treatment (with a diagnosis) or triggering an offer of treatment (with a treatment indication) can be based on the principles described above.

The diagnostic criteria for diabetic albuminuria and nephropathy

Diagnostic criteria are based on categorical variables (e.g., a sample collected on rising) or dichotomised results of discrete or continuous variables (e.g., an AER above 40mcg/min). A diagnostic criterion is usually made up of a combination of such variables or findings. This combination can be assembled using a logical process that begins with one 'lead' finding, which is a manageably short list of differential diagnoses that account for nearly all those to whom the list applies [12]. In this case, the lead of 'albuminuria' is an average AER of three measurement results that are least 20mcg/min. To assemble the combination of diagnostic findings for DA the other known causes of albuminuria must be 'excluded'. In the case of diabetic nephropathy, the 'lead would be an average AER above 200mcg/min from 3 separate measurements and an AER rise of at least 30% from baseline. The positioning of this threshold could be reviewed based on a decision process when offering the interventions suggested by the diagnosis of nephropathy.

A combination of diagnostic findings is assembled by looking for findings that cannot occur by convention in the lead's other possibilities (e.g., an HbA1c consistently above 48mmol/l cannot occur by convention in 'non-diabetic albuminuria'.) Similarly, heavy exercise or orthostatic albuminuria is regarded as excluded if the urine sample is collected on rising from sleep. A urinary tract infection or nephritis is excluded by an MSU showing no red or white cells and no bacterial growth. We must also include in our list the possibility of some other unknown diagnosis not named. As each item of evidence excluding other possibilities comes in, the probabilities of those not excluded rises until only DA or something not in the list remains. There is a theorem based on an assumption of conditional dependence that models the changing probabilities during this thought process [12].

If the reasoning takes place with diagnostic criteria, then a simplification can be used. For example, if the probabilities of the differential diagnosis albuminuria were DA (0.5), non-diabetic albuminuria (0.2), heavy exercise (0.1), UTI (0.1), prolonged standing (0.08), Nephritis (0.01), something not in this list (0.01), these would add to 1. The estimated probability of DA if none of the possibilities were excluded can be found by subtracting the probability of each other diagnosis from one:
1- 0.2-0.1—0.1-0.08-0.01-0.01 = 0.5. If the urine specimen was taken after working a nightshift (i.e., heavy exercise and prolonged standing could not be excluded so that their probabilities remained 0.05) then the estimated probability of DA would be 1-0-0.1—0-0.08-0-0.01 = 0.81. If the other possible diagnoses could not be excluded by as they did not involve diagnostic criteria, then a more sophisticated expression would have to be used (12, 13, 14). If all the possibilities were excluded so that the probabilities of those excluded were zero (except for something not on the list) then the estimated probability of DA would be 1-0-0-0-0-0-0.01 = 0.99. Ideally, the unlisted possibilities should be identified and excluded, allowing the combination of findings to be a 'sufficient' criterion for DA. However, we might have to accept that the diagnostic criterion might include 1% of other unknown diseases and result in a degree of over-diagnosis until the unknown diagnoses can be identified.

Confirming a diagnosis is to assume that a disease process is present (N.B. 'assuming' does not mean that it is present). A choice of different combinations of findings can be used to confirm DA so that the above combination may be one of several 'sufficient' criteria for the diagnose of DA (another combination might use the albumin creatinine ratio (ACR) instead of the AER). These criteria may also identify slightly different populations with the diagnosis of DA. A definitive criterion identifies all those and only those with a diagnosis of DA (i.e., the diagnosis that is an assumption but not the 'true disease' of DA with certainty), so that would have to include all the individual sufficient criteria.

The purpose of a diagnosis and its criteria is to identify patients who are likely to benefit from interventions suggested by that diagnosis. Its definitive criteria should be designed as far as possible to avoid omitting those who benefit but also to avoid including those with no prospect of benefit. Diagnoses may then be subdivided by looking for treatment indication criteria that increase the probability of benefit based on a meaningful difference between the probability of an outcome on treatment and control. The most obvious way of doing this is based on the severity of the condition as described here. It is also possible to compare the effect of using different variables (e.g., the AER and ACR) to see what effect they have on the steepness of curves as shown in Figure 2 and their ability to create high and low outcome probabilities to improve decision making.

If all the possible 'sufficient' diagnostic combinations that confirm DA are taken together then they will identify all those and only those with the diagnosis. However, this does not mean that they will identify all those with the underlying disease as some in the population may not have been tested for diabetes mellitus or albuminuria. In this sense, it is not possible to have a definitive criterion that identifies all those and only those with a disease. It is only possible to do so for a diagnosis, which is created by doctors to apply medical knowledge to individual patients.

Risk ratio or risk reduction as a measure of treatment efficacy

It is widely assumed that the risk ratio or risk reduction as observed from a RCT is constant for different baseline risks and therefore can be used to estimate the absolute risk reduction for a range of risks [5, 13]. Others argue that it is the odds ratio is a better for this purpose [14]. Figure 4 shows that in this study, neither the risk ratio nor the odds ratio is constant for all values of AER and baseline risk of nephropathy conditional on the control intervention. This was observed because a calibrated logistic regression function was fitted separately to the control and treatment data.

In Figure 4, the risk ratio is shown to be nearer 1 (i.e., the treatment effect is weaker) at high and low values of the AER and the risk ratio is lower (i.e., the treatment is more effective) in the mid AER range around 110mcg/min. This curve (as opposed to a straight horizontal line) represents treatment heterogeneity based on diagnostic severity. The treatment effect is also dose dependent of course, the risk ratios being lower and the treatment having more effect on the higher dose of Irbesartan 300mg daily than on Irbesartan 150mg daily. There is also patient heterogeneity due to the range of diagnostic severity represented by the AER results between 20 and 200mcg/min for the people in the trial. If the results are only recoded for the overall range as in Table1, this heterogeneity is hidden of course. Figure 3 shows the combined effect of treatment and patient heterogeneity.

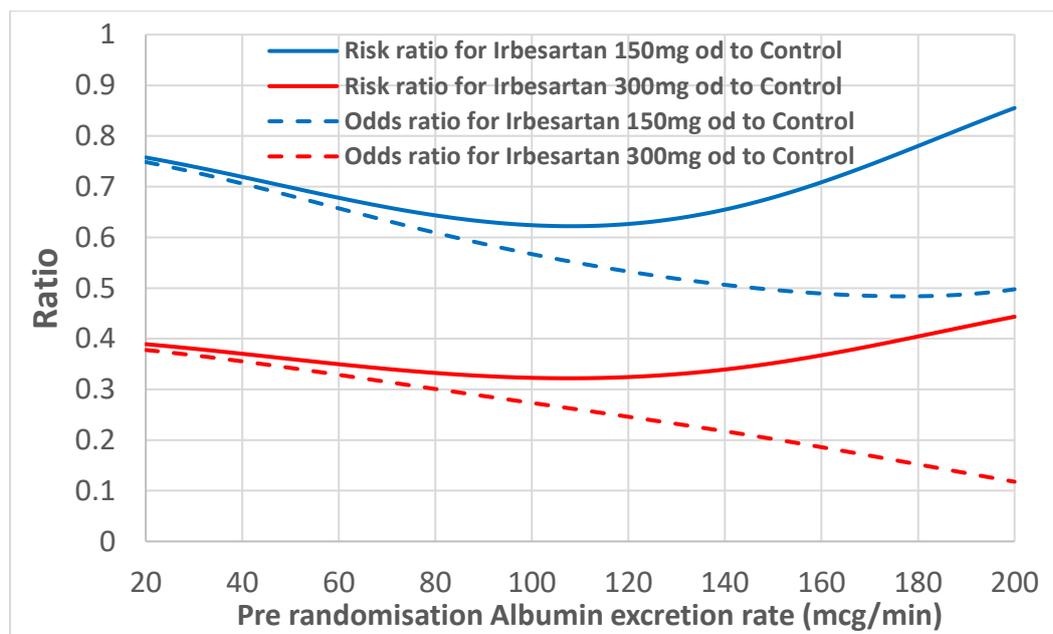

Figure 4: The change in risk ratio and odds ratio for different values of the AER after Irbesartan 150 and 300mg daily

Figure 5 shows the probability of nephropathy on treatment with Irbesartan 150 and 300mg daily when estimated using the calibrated logistic regression compared to assuming a constant risk ratio and constant odds ratio. The curves are almost superimposed between an AER of 20 and 90mcg/min. Therefore, any of the three methods would suffice to assess the performance of tests for establishing a threshold at low probabilities (e.g., <0.1) for diagnosis or offering treatment. For higher probabilities, the curves for odds ratio and odds ratio either underestimate or overestimate the calibrated result indicating that they may not be consistent with the raw data in some ranges as shown in Table 2. If it is important for all probabilities to be consistent with the data, it would be sensible to regard the assumption of a constant risk or odds ratios as provisional, and to calibrate such preliminary probabilities.

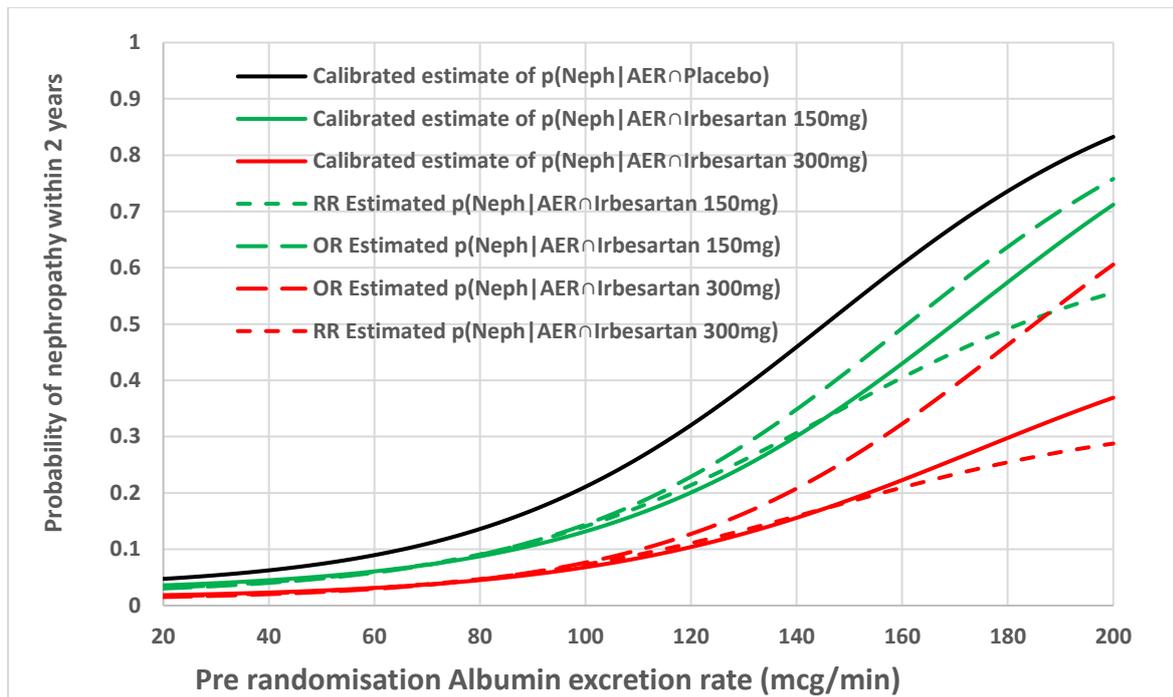

Figure 5: The probability of nephropathy after irbesartan 150 and 300mg daily estimated using calibrated logistic regression, risk ratio and odds ratio.

Multiple risk factors and causal inference

The RCT was conducted on subjects with well controlled blood pressures (an average of 143/83mmHg) and well controlled diabetes (an average HbA1c of 7.1% = 54mmol/mol). Because of this, when the AER was low, the risk of nephropathy due to the other risk factors was also low. However, if the patient referred to earlier who had an AER of 40mcg/min also had a very high blood pressure and very high HbA1c, then the risk of nephropathy on control intervention might be 0.2 (instead of 0.062). From Figure 3, the risk ratio at 40mcgm/min is 0.370 and from Figure 2, the risk difference at 40mcg/min is 0.062-0.023 = 0.039. If we apply the risk ratio (AKA the relative risk), then the new risk after treatment with Irbesartan 300mg daily would be 0.2x0.37 = 0.074, representing a dramatic risk reduction of 0.126. However, if we apply the risk difference (AKA the absolute risk reduction) then the new risk of nephropathy would be 0.2-0.039 = 0.161, which is a more modest risk reduction of 0.039.

If we apply the risk ratio of 0.370, then we are assuming that Irbesartan reduces the blood pressure and HbA1c in addition to the AER to produce a much lower risk of nephropathy. However, if we apply the risk difference of 0.039, then we assume that Irbesartan reduces the AER alone to lower the risk of nephropathy. In this case however, the Irbesartan would also reduce the systemic blood pressure perhaps via a different mechanism. This means that we should be applying another risk difference for this other treatment effect. We would not expect Irbesartan to reduce the HbA1c of course.

In the discipline of 'causal inference', using the risk ratio is known as applying a treatment effect on the 'multiplicative scale' and using the risk difference is known as applying a treatment effect on the

'additive scale' [15]. In estimating the effect of statins on vascular risk, the risk reduction for an individual is usually calculated on the 'multiplicative scale' (e.g., in the Mayo clinic statin choice decision aid [16]), which is perhaps questionable because its multivariable risk calculation is based on linear regression and assumes 'additivity'. It should be noted that the curves in Figure 2 are not linear so that a multivariable risk calculation for nephropathy that uses the AER should not assume linearity and that a different assumption should be made in the calculation [17].

Other ways of assessing diagnostic tests

Diagnostic test accuracy involves comparing a new test with an established test, the latter being described as the reference or gold standard. The new test is assessed by first dichotomising its numerical results (e.g., into those within or outside the 'normal range' of the test population) and then assessing its sensitivity and specificity with respect to a reference standard. A new test that performs well in this way is regarded as promising for use in clinical practice. Its clinical performance can then be assessed by randomising subjects to different groups so that use of the new test in one group is compared to the use of an established test in another group to determine which test produces the best outcome in terms of benefit. The problem with this is that such a test is not assessed currently in a way that the probabilities can be estimated conditional on each numerical result, and this used to establish thresholds for diagnosis and offering treatment that avoid over-diagnosis and under-diagnosis. This point appears to be supported by a suggestion that the evaluation of tests for use in diagnosis and treatment selection should be assessed with RCTs as opposed to being a part of an assessment of test accuracy [18].

Practicalities of general application

The raw data required to implement this approach may be available from past RCTs as was the case for the IRMA2 trial used in this example. Failing this, new studies would have to be done perhaps with novel designs that only involve randomisation of subjects with lower risks of outcomes (around the expected positioning of thresholds at lowish probabilities) but requiring larger numbers of patients. A Bayesian approach could also be employed by estimating the curves such as those in Figures 1 to 5 subjectively by several doctors with experience of the topic and then updating the curves when data become available from studies.

Conclusion

To avoid over-diagnosis and over-treatment, tests need to be assessed so that the probabilities of outcomes can be estimated conditional on each numerical test result, presented graphically for use during shared decision making, and used to establish pragmatic thresholds for confirming diagnoses and offering treatment.

Acknowledgements

I am grateful for the support of Sanofi-Synthelabo and Bristol-Myers Squibb, particularly Dr Gérald Frangin and Dr Joanne Clarke, and the other investigators in numerous countries who also participated in the study that provided the data used in this paper. I am also grateful for those who made suggestions and participated in the discussions about this work at a presentation and seminar at the Preventing Overdiagnosis 2023 International Conference in Copenhagen during August 2023.

Competing interests

I have no competing interests to declare.